# CAPITAL STRUCTURE IN U.S., A QUANTILE REGRESSION APPROACH WITH MACROECONOMIC IMPACTS

Andreas Kaloudis[1], Dimitrios Tsolis[2]

[1] PhD Candidate, Cultural Heritage and New Technologies Department, University of Patras, antreaskaloudis25@gmail.com.

[2] Assistant Professor, Cultural Heritage and New Technologies Department, University of Patras, dtsolis@upatras.gr.

**Abstract**

The major perspective of this paper is to provide more evidence into the empirical determinants of capital structure adjustment in different macroeconomics states by focusing and discussing the relative importance of firm-specific and macroeconomic characteristics from an alternative scope in U.S. This study extends the empirical research on the topic of capital structure by focusing on a quantile regression method to investigate the behavior of firm-specific characteristics and macroeconomic variables across all quantiles of distribution of leverage (total debt, long-terms debt and short-terms debt). Thus, based on a partial adjustment model, we find that long-term and short-term debt ratios varying regarding their partial adjustment speeds; the short-term debt raises up while the long-term debt ratio slows down for same periods.

**INTRODUCTION**

General (and economic) managers with a view to maximize profits must make two fundamental choices: the choices of investment and capital structure. Myers (2001) supported that there is no single theory that could explain the lending leverage of companies, and there is no reason to expect one. However, many academics have attempted to correlate variables with the capital structure, with the prevailing theories being the pecking-order theory, the trade-off theory, the theory of representation costs and the market-timing theory. The early studies, yet, had no obvious effect. For example, the trade-off theory supports that there is a positive correlation between debt leverage and profits. In contrast to their anticipations, Rajan and Zingales (1995) underpin a negative correlation between leverage and profits. Lemmon and Zender (2010) identified a positive relation between the lending leverage and the market value index per book value. On the other hand, Sinan (2010) found that the lending leverage is negatively correlated with the market value index per book value. Myers's point of view is that there is no single theory that can explicate the choice of capital structure from companies and that the factors fluctuated depending on the country and the period investigated, strengthened by these contradictory results. In general, the applicability of the capital structure theories enlists majestically on the research agenda, and a growing number of country-specific studies have extended the investigations.

In this context, Stephen D. Prowse (1990) examined the agency problem between debtholders and shareholders of Japanese and U.S. firms. Leverage ratios of U.S. firms are negatively related to the enterprises potential engagement in risky investments whereas Japanese debt ratios show no such relation. Evidence is consistent with the concept that agency problem is alleviated to a superior degree in Japan than in the U.S. Raghuram G. Rajan and Luigi Zingales(1996) analyzed the determinants of capital structure by investigating decisions of public firms in the major industrialized countries and provide evidence that in an aggregate level, firm leverage in U.S. is similarly correlated in other G-7 countries in a cross-section sample. Moreover, estimations indicate that leverage in U.S. firms is positively related to asset tangibility and firm size, and negatively related to market-to-book ratio and profitability, followed by more recent international evidence. (Booth. (2001), and De Jong. (2008)). As described by John K. Graham (1996), by using annual data from more than 10,000 firms from 1980 to 1992, it is tested whether the progressive use of debt is correlated positively to simulated firm-specific marginal tax rates that account for net operating losses, alternative minimum tax and investment tax credits. The author provides evidence which point out that high-tax-rate firms issue less low-tax-rate counterparts than their debt. In his findings, John K. Wald (1999), in an empirical study, examined the factors which are correlated with the capital structure in United States, United Kingdom, Germany, Japan and France and the results illustrate similarities in leverage among all countries. However, differences appear in the correlation between long-term debt to asset ratios and profitability, size and growth. Specifically, the author found a negative correlation between leverage and profitability as prior researches of Raghuram G. Rajan and Luigi Zingales(1996), through a regression analysis in a total sample of 3,300 firms in United States. The findings of this study suggest links between varying choices in capital structure across countries as well as legal and institutional differences. Shuenn-Ren Cheng, Cheng-Yi Shiub.(2006) using a sample of firms across 45 countries including U.S., found that investor protection plays an important role in the determinants of capital structure: firms in countries with better creditor protection have higher leverage, while firms in countries where shareholder rights are better protected use more equity funds. Rongbing Huang and Jay R. Ritter (2009) examined time-series patterns of external financing decisions and illustrated that when the cost of equity capital is low, publicly traded U.S. firms fund a larger quantum of their financing deficit with extraneous equity. The historical values of the cost of equity capital have long-lasting impact on capital structure of the firms through their effect on firms' historical financing decisions. The authors also present an alternative and innovated econometric technique to deal with biases regarding estimates of the speed of adjustment toward target leverage. The research provides evidence that firms adjust toward target leverage at a passable speed, with a half-life of 3.7 years for book leverage, similarly after controlling of the conventional determinants of firm fixed effects and capital structure. Consequently, with the market timing theory, Rongbing Huang and Jay R. Ritter found that when the expected ERP is lower, firms fund a larger proportion of their financing deficit with net external equity. As described by Reint Gropp and Florian Heider (2010) in a sample of large U.S. and European Banks found that unobserved time-invariant bank fixed-effects are eventually the most important determinants of banks' capital structures and that banks' leverage converges to bank specific, time-invariant targets. However, there is a significant number of references in literature, focused principally on samples in United States (Jacelly Cespedes, Maximilliano Gonzalez, Carlos A. Molina (2010), Douglas O.Cook, Tian Tang (2010), Özde Öztekin, Mark J. Flannery (2011), Amarjit Gill, Nahum Biger, Neil Mathur(2011) ). Jacelly Cespedes, Maximilliano Gonzalez, Carlos A. Molina (2010) evaluated the capital structure determinants of Latin American firms by using a sample of seven countries. Latin American firms have high ownership concentration, which is a fact that creates an ideal setting to study how ownership concentration explains firms' capital structure. This study finds a positive relation between leverage and ownership concentration. Furthermore, the study indicates a positive relation between leverage and growth. Thus, firms that are larger have more tangible assets whereas less profitable firms are more leveraged.

A key element of this paper is the utilization of limited and unlimited relapses to distinguish the firms impact from the firm-specific and macroeconomic variables. The outcomes demonstrate that due to firm-specific characteristics varieties occur.

Douglas O.Cook, Tian Tang (2010) by using two dynamic partial adjustment capital structure models estimated the impact of several macroeconomic factors on the speed of capital structure adjustment toward target leverage and found evidence which confirm that firms adjust their leverage toward target faster in optimal macroeconomic states compared to weak macroeconomic states. Hall et al. (2006) demonstrated that long-term obligations are connected decidedly to the resource structure and firm size and contrarily to age; fleeting obligations are connected adversely to benefit, resource structure, size and age and emphatically to development. A huge variety crosswise over enterprises was found in a large portion of the illustrative variables. Özde Öztekin, Mark J. Flannery (2011), as Douglas O.Cook, Tian Tang (2010), compare firms' capital structure adjustments across countries and investigate whether institutional differences are appropriate to explain the variance in estimated adjustment speeds. Authors in consistence with the dynamic Trade-off theory, found that legal and financial traditions significantly correlate with firm adjustment speeds. Sinan (2010) demonstrated that there is a positive correlation between leverage and profitability. This is because, when financial leverage is used, changes in earnings before interest and taxes bring greater changes in profit before disposal per share. High leverage simply means that small changes in sales incur disproportionately larger changes in the operating profit and vice versa. More recently, Amarjit Gill, Nahum Biger, Neil Mathur (2011) selected a sample of 272 American firms listed on New York Stock Exchange for a three-years period, from 2005 to 2007. Empirical results show positive relationship between total debt to total assets and profitability in the service industry, in contrast with Raghuram G. Rajan and Luigi Zingales(1996) findings. On the contrary, in the manufacturing industry the results indicate three positive relationships between long-term debt to total assets and profitability, short-term debt to total assets and profitability, and total debt to total assets and profitability. In the findings of Mokhova and Zinecker (2013), the relationship examination between the capital structure and the sovereign FICO scores demonstrates the distinctions in appraisal valuation by rating firms, which can be clarified by various financial variables and their weights in the connected default likelihood models. The quality of the connection between the capital structure and the FICO assessments additionally relies on the measures of the capital structure and the nation's specifics. Koksal and Orman (2015) argued that the trade-off theory gives a superior portrayal of the capital structures of every single firm to the pecking-order theory. In addition, the trade-off theory appears to be especially appropriate for comprehending the financing decisions of extensive private firms in the non-producing sector and when the financial environment is generally steady. The study's conclusion is that the trade-off theory is a superior system to the pecking-order theory.

However, some authors examined the capital structure determinants with the method of quantile regression in order to study the effect of leverage in firms across different quantiles. Bassam Fattouh, Laurence Harris and Pasquale Scaramozzino (2008) demonstrated that, by presuming upon the distribution of leverage, conditional quantile regression methods generate new information about the most appropriate choice of leverage ratio in a UK companies sample finding that not only the estimated effect of the regressors is different at different quantiles of the distribution, but also that the effect of a variable changes sign between low leveraged and high leveraged firms. Dimitris Margaritis and Maria Psillaki (2010) investigated the relationship between leverage and firm efficiency. Authors considered both the effect of leverage on firm performance as well as the reverse causality relationship and tested the attitude of leverage across different quantiles using a sample of 12,240 New Zealand firms. They found evidence supporting the theoretical predictions of the Jensen and Meckling (1976) agency cost model using quantile regression analysis.

We try to determine which of capital structure theories (trade-off theory, agency cost theory and pecking-order theory) is the most suitable for explaining empirically the capital structure of US firms. Therefore, the tasks that are going to be carried out in this study are the analysis of the variables for each of the theories and the identification of the one that best explains the index of long-term debt leverage, short-term debt leverage and total debt leverage. In this paper the applicability of capital structure theories in the US economy is examined by a panel data empirical model that uses their key variables. Using panel models of random or fixed effects, it is possible to control the implications of companies' non-observable individual effects on the estimated parameters. The period of analysis spans from 1970 to 2014, incorporating different market phases and various stock market crashes and booms. The years before the financial crisis were characterized by an excessive accumulation of exposures of US firms in relation to their own funds (leverage).

We find a negative correlation between leverage and the profitability and the firm's size. Our results imply that firms source finance in a manner consistent with Myers's (1984) pecking-order theory. Furthermore, we find a positive correlation between tangible assets and leverage but a negative correlation between liquidity and leverage, which is also consistent with Myers's (1984) pecking-order theory. The capital requirements for covering risks are essential to ensure sufficient own funds to cover unexpected losses. However, the crisis has shown that these requirements alone are not sufficient to prevent U.S. firms from taking excessive and unsustainable leverage risk. This means that U.S. firms do not increase or reduce their debt with their investment opportunities, as shown in Kester (1986)'s study. Additionally, we observe a negative correlation appears between leverage and firm size. This means that U.S. firms do not increase their external debt with their investment opportunities but also do not direct internal financing and additional borrowing, but that exposes them to large systemic risk. It is also demonstrated that deficits are financed with debt, and similar results were found by Shyam-Sunder and Myers (1999) for US companies.

None of the above studies combine firm specificities with adjustment speed in capital structure determination and research the issue of macroeconomic variables vs firm-specific characteristics in different economic states with a quantile regression approach. The main purpose of this paper is to contribute to the relative importance macroeconomic variables vs firm-specific characteristics when the macroeconomic conditions change.

To the best of our knowledge, this paper is the first to examine the applicability of capital structure theories to US firms with an extensive data set in this period. Our study differs from previous studies in literature in many ways. We propose to make a number of contributions to the literature.

This paper, thus contributes on the recent of capital structure dynamic determination in the following ways: we examine which variables affect the leverage of a very large data of the US firms for the first time with the largest number of data, the research is expanded to cover almost all business categories. The result is the increment of the sample to 28793 companies and 340865 observations, which is the largest sample size in the research literature concerning the economy of the United States (US), the period of analysis is the longest so far in the research literature; in addition, the previous studies are scattered and the approach methodology is quantile regression in a fixed effects model in different macroeconomic states.

The remainder of this paper proceeds as follows. Section 2 describes the data and the methodological approach, the methodology set and some preliminary statistics. Section 3 presents and the Model. Section 4 provides the empirical results.

**METHODOLOGY AND DATA**

For the beginning we examine if our panel data model is a fixed or random effects model. Through Hausman test we found that our model is a fixed effect model. Panel data models allow us to control the implications of companies' non-observable individual effects on the estimated parameters. For the best of our knowledge panel data are the most suitable to examine a dynamic phenomenon that changes cross time in comparison with time series or cross-section data which neither express dynamic relations nor produced estimates are highly accurate due to the multicollinearity existence. Furthermore, panel data provide us estimates of raised accuracy while they used more than the double number of total observations that is used in both assessment with the times series or cross section data. Finally, with panel data, there is the possibility to control the invariable elements that change between firms but are stable over time. To decide the most appropriate model between fixed or random effects model, through Hausman (1978) test, the fixed-effects model is the most appropriate for our research.

**THE MODEL**

Following the rationale of Cook and Tang (2010) and Oztekin and Flannery (2012),we use a partial adjustment model which assumes that the target debt ratio $DR *_{i,t}$ from firm $i$ at time $t$, is given by:

$$DR *_{i,t} = a * + a *_i + \beta * X_{i,t-1} + \gamma * M_{t-1}, \quad i = 1,\ldots\ldots,N, \quad t = 2,\ldots\ldots,T_i$$

Where $a *$ is the constant term, $a *_i$ is the unobserved heterogeneity of firm $i$, $X = (X_1, \ldots, X_K)'$ and $M = (M_1, \ldots, M_J)'$ are (column) vectors of firm specific and macroeconomic variables respectively, $\beta * = (\beta *_1, \ldots, \beta *_K)$ is the (row) coefficient vector of firm-specific variables and $\gamma * = (\gamma *_1, \ldots, \gamma *_j)$ the (row) coefficient of the macroeconomic variables. The debt ratio $DR_{i,t}$ adjust to its target according to the rule:

$$DR_{i,t} - DR_{i,t-1} = \delta * (DR *_{i,t} - DR_{i,t-1}) + \varepsilon_{i,t}$$

**Where** $\delta *$ is the speed of adjustment and $\varepsilon_{i,t}$ is the error term.

# THE ESTIMATION METHODOOGY

## The Quantile regression in Panel Data approach

Our approach is based on quantile regressions, which estimate the effect of explanatory variables on the dependent variable at different points of the dependent variable's conditional distribution.

Quantile regressions were originally presented as a 'robust' regression method which permits for estimation where the typical hypothesis of normality of the error term might not be strictly satisfied (Koenker and Bassett, 1978);

This method has also been used to estimate models with censoring (Powell, 1984, 1986; Buchinsky, 1994, 1995).

Recently, quantile regressions have been used simply to get evidence about points in the distribution of the dependent variable further than the conditional mean (Buchinsky, 1994, 1995; Eide and Showalter, 1997). We use quantile regressions to observe whether the effects of factors Is differentiated across the 'quantiles' in the conditional distribution of dependent variable

As described by Koenker and Bassett (1978), the estimation is done by minimizing

$$\underset{\beta \in R^K}{Min} \sum_{t \in \{t: y_t \geq x_t\beta\}} \theta |y_t - x_t\beta| + \sum_{t \in \{t: y_t < x_t\beta\}} (1-\theta)|y_t - x_t\beta|$$

where $y_t$, is the dependent variable, $x_t$ is the k by 1 vector of explanatory variables, β is the coefficient vector and Θ is the estimated quantile. The coefficient vector b will differ depending on the particular quantile being estimated.

## Interpretation of Quantile Regression Estimation

Since the $\theta_{th}$ conditional quantile of y given x is given by

$$Quant_\theta(y_i|x_i) = \acute{x}_\iota \beta_\theta,$$

its estimate is given by, $Quant_\theta(y_i|x_i) = x'_i \hat{\beta}_\theta$.

As one rises θ continuously from 0 to 1, one drops entire conditional distribution of y, conditional on x. In practice, given that any data set contains only limited number of observations, only a finite number of quantile estimates will be statistically distinct, although this number can be quite large.

Consider the partial derivative of the conditional quantile of y with respect to one of the regressors, say j, namely

$$\partial Quant_\theta(y_i|x_j)\acute{x}_\iota/\partial x_{i,j}$$

This derivative is to be interpreted as the marginal change in the $\theta_{th}$ conditional quantile due to marginal change in the $j_{th}$ element of x.

If $x$ includes $K$ distinct variables, then this derivative is given by $\beta_{\theta j}$, the coefficient on the $j_{th}$ variable. One should be careful in interpreting this result. It does not suggest that a person who happens to be in the $\theta_{th}$ quantile of the conditional distribution will also find himself/herself at the same quantile had his/her x changed.

**Quantile Regression in Panel Data**

Like in the case of the conditional mean, for the estimation of the conditional quantile equation we use different methods when employing panel data than those used for cross-sectional or time series data. Similar to conditional the mean, we use the FE model for quantile regression for the estimation of the conditional quantile. Since the FE model is the less restrictive model and we do not have to assume the absence of correlation between regressors and individual effects, we consider that FE is the most appropriate model for this application. Additionally, there is no generally agreed upon notion regarding random effects for quantile regression applications. On the other hand, there exist penalized methods (Koenker, 2004) that are used for longitudinal data that improve the efficiency of the FE model. However, they mostly suited for cases where there are just few observations (usually less than 5) of one of the "two-way" effects, time or individuals. Nevertheless, no theoretical background has been developed yet to use "two-way" penalized methods. Therefore, even if it was more appropriate, it would be infeasible to use. In cases where we needed to improve the FE model's efficiency we could estimate a model where quantile regressions are estimated simultaneously by imposing identical individual effects for every quantile and minimizing the equation (Koenker, 2004)

We deliberate a panel QR model with individual effects

1. $y_{it} = \alpha_{i0}(\tau) + x'_{it}\beta_0(\tau) + Z'_{it}\gamma_0(\tau) + e_{it}(\tau), i = 1, \ldots, N, t = 1, \ldots, T.$

$y_{it}$ is the scalar response for the i-th individual at t-th time period, $x_{it}$ and $z_{it}$ are $dx \times 1$ and $dz \times 1$ vectors of covariates, $\alpha_{i0}(\tau)$ is the fixed effect for $i_{th}$ individual.

For a brace of observations $\{(y_{it}, x_{it}, z_{it}), t \geq 1\}$, we permit time necessity in a given individual, but accept independence across individuals.

We assume that the $\tau_{th}$ quantile of the error $e_{it}(\tau) = 0$.

Under this normalization situation, the $\tau_{th}$ conditional quantile of the response given covariates is written as

$Qy_{it}(\tau|x_{it}, z_{it}) = \alpha_{i0}(\tau) + x_{0it}\beta_0(\tau) + z_{0it}\gamma_0(\tau).$

The quantile specific error $e_{it}(\tau)$ illustrates the distance between the response $y_{it}$ and its $\tau_{th}$ conditional quantile. We allow the marginal densities of $e_{it}$ to be distinct at different time periods. Let $ft(e)$ be the marginal density of $e_{it}$ and $f(e) = T^{-1}\sum_{t=1}^{T} ft(e)$.

We use a vector notation α(τ) to signify a set of fixed effects $(\alpha 1(\tau), \ldots, \alpha N(\tau))$.

The QR coefficients are estimated by

2. $\left(\hat{\alpha}(\tau), \hat{\beta}(\tau), \hat{\gamma}(\tau)\right) = \underset{a(\tau),\beta(\tau),\gamma(\tau)}{\arg\min} \sum_{i=1}^{N}\sum_{t=1}^{T} \rho_\tau\left(y_{it} - a_i(\tau) - x'_{it}\beta(\tau) - Z'_{it}\gamma(\tau)\right).$

$\rho_\tau(u) = \tau - |(u \leq 0)|$, u is the check function as in Koenker and Bassett(1978).

**The model**

$$DR_{i,t} = a + a_i + a^c c_t + \delta DR_{i,t-1} + \delta^c DR_{i,t-1}C_t + \beta X_{i,t-1} + \beta^c X_{i,t-1}C_t + \gamma M_{t-1} + \gamma^c M_{t-1}C_t + e_{i,t}$$

Where $a$ is the constant term, $a_i$ is the unobserved heterogeneity of firm $i$, $X = (X_1, \ldots, X_K)'$ and $M = (M_1, \ldots, M_J)'$ are (column) vectors of firm specific and macroeconomic variables respectively, $\beta* = (\beta*_1, \ldots, \beta*_K)$ is the (row) coefficient vector of firm-specific variables and $\gamma* = (\gamma*_1, \ldots, \gamma*_j)$ the (row) coefficient of the macroeconomic variables

Cook and Tag (2010) examine the above model using panel data in a sample of US firms over the period 1977-2006 and expand further the model by including a dummy variable for the good and bad states of the economy and interacting it with the lagged debt ratio.

We set dummy variable for good(growth) and bad(recessionary) macroeconomic states, and we follow the model of Cook and Tang (2010) in a fixed effect quantile regression approach in order to capture the fluctuations of the regressors across all over the distribution of total, short-terms and long-terms debt.

Where $c_t$ takes value 1 if gdp growth rate is negative and 0 otherwise. So the effects of the lagged debt ratio, the lagged firm-specific variables and the lagged macroeconomic variables on the debt ratio are given by $\delta$, $\beta$, and $\gamma$ in the good state and $\delta + \delta^c, \beta + \beta^c, \gamma + \gamma^c$ in the bad state

**The variables**

Proxies for leverage

The leverage variable is differentiated between three types which are the total debt ratio(TDR), the long-term debt ratio (LDTR) and the short-term debt ratio (STDR) in order to capture all angles from prior literature. This categorization of leverage allows us to investigate the influences of debt maturity structure across macroeconomic states.

The calculation of ratios participated by following the standard practice by measuring debt ratios as the book value of interest-bearing debt to total assets. Therefore, TDR is firm's short-term plus long-term book value of debt divided from total assets.

**LTDR** is long-term debt divided from total assets and STDR is the short-term debt divided from total assets.

**Firm-specific factors**

The firm-specific factors are represented by the standard set of capital structure determinants.

**Size (SIZE)** expected to be positively correlated with debt levels. Larger firms may be able to reduce the transaction costs associated with long-term debt issuance. Public corporate debt usually trades in large blocks relative to the size of an equity trade, and most issues are at least 100 million dollars in face value to provide liquidity. Larger firms may also have a better chance of attracting a debt analyst to provide information to the public about the issue. Another possibility is that larger firms have more dilute ownership, and thus less control over managers. Managers may then issue less debt to decrease the risks of bankruptcy that involve personal loss (see Friend and Lang (1988) and Friend and Hasbrouck (1988)). Marsh's (1982) survey concludes that large firms more often choose long-term debt while small firms choose short-term debt. Size is measured as the natural logarithm of total sales.

**Asset structure (AS)** is expected to be positive or negative correlated with debt levels. The positive sign reflects the guarantees that banks require from companies in order to grand loans.

Regarding to Myers (1977) companies with **growth** potential are assumed as riskier and in the future, they will tend to have lower leverage. On the contrary, it is more likely that firms with high growth opportunities exhaust internal funds and seek external financing (Michaelas et al., 1999). As a consequence, growth (GR) and leverage relation can be either negative or positive, with GR being measured as the annual rate of change in sales.

In view of the pecking order theory, SMEs financing decisions follow in general a hierarchy, preferring debt over equity and internal over external financing (Michaelas et al., 1999, Daskalakis and Psillaki (2008), Psillaki and Daskalakis, 2009). Thus, it is expected that **profitability** (PR) should be negatively related to debt and be measured as earnings before interest and taxes to total assets.

It should be noted that tax reflections are of little attention for SMEs (Pettit and Singer, 1985), since these firms are less likely to generate high profits and as a result less likely to use debt or non-debt items for tax shields. Nevertheless, considering the given higher levels of difficulty for small firms in order to access debt financing, we can deduce that the use of **non-debt tax shields** (NDTS) could be viewed as their main alternative for reducing any tax burdens. Hence, it is easily inferred that non-debt tax shields will be negatively related or not related to debt (Titman and Wessels, 1988). NDTS is calculated as the ratio of total depreciation expenses to total assets.

On the other hand, **riskier** firms (RISK) are to confront higher levels of difficulty in accessing debt financing (DeAngelo and Masulis, 1980; Titman and Wessels,1988). Consequently, it is expected a negative relationship between leverage and risk. As far as RISK calculation is concerned, we consider a three-year rolling window of the earnings standard deviation before interest and taxes.

Concerning the significance of **trade credit** (NTCS) as a source of short-term financing, particularly for SMEs, it is derived that it is well-documented (Ng et al., 1999; Asselbergh, 2002; Guariglia and Matut, 2006). Two alternative hypotheses exist in the literature regarding the use of trade credit: the first one is the substitution hypothesis and the other one is the complementarity hypothesis. Moreover, evidence shows that trade credit acts as complement, during times of tight money, rather than substitute to bank credit, providing support for the redistribution effect (Love et al., 2007; Casey and OToole, 2014; Psillaki and Eleftheriou, 2015). Hence, we anticipate either a negative or a positive relationship of trade credit before the crisis that follows the substitution or the complementarity hypothesis respectively, and during the crisis a positive relationship. As for NTCS, the net-trade approach of Love et al. (2007) is followed; firstly, trade payables are subtracted from trade receivables and then divided by total sales.

It is also expected that cash-rich firms (CASHTA) will have lower debt for two specific reasons; firstly, due to the fact that risky firms will definitely try to accumulate cash with a view to avoid in the future under-investment issues and secondly, cash-rich companies will choose internal financing as clarified in the pecking theory context. In an effort to estimate CASHTA, we compute it as cash to total assets.

Finally, we consider the firm-specific time varying interest burden (FINEXP) for the purpose of capturing the fact that each firm faces different interest burden, and we expect a negative relationship between the lagged value of financial burden and leverage. FINEXP is calculated as the financial expenses to sales ratio.

**Macroeconomic factors**

An extensive literature is provided by Mokhova and Zinecker (2014) upon the specific issue of the effect of various macroeconomic factors on corporate capital structure. Taken for granted that commercial banks constitute the most common source of external financing for SMEs (Colombo and Grilli, 2007; De Bettignies and Brander, 2007) credit supply (CRED) is used as one of our macroeconomic variables, anticipating a positive relationship between credit supply and leverage; we expect either expansion during the growth stage or contraction during the recessionary stage. We estimate CRED as the annual growth rate of total credit expansion to enterprises and households.

Another extensively investigated macroeconomic factor is the inflation rate (INFL). However, contradictory evidence exists concerning the effect of inflation on capital structure. In the context of literature, Bastos et al. (2009) find no effect of inflation on leverage, while Frank and Goyal (2009) detect a positive relationship between market leverage and inflation, yet no relationship on book leverage. On the other hand, Hanousek and Shyamshur (2011) discover that inflation generally has a positive influence on leverage, this effect however turns unimportant for certain specifications of their model. INFL is referred to the annual rate of change of the CPI index.

Furthermore, the interest rate on which companies borrow is another significant macroeconomic factor that represents the cost of debt. The relationship between interest rates, macroeconomic conditions and firm's leverage has been thoroughly examined by researchers (e.g. Karpavicius and Yu, 2017; Halling et al., 2016; Baum et al., 2009; Frank and Goyal, 2004; Korajczyk and Levy, 2003) and it is derived that the empirical evidence on the relationship between interest rates and firm's leverage is varied. Notwithstanding the fact that most of the researchers do no approach the relation from the company's perspective, namely that companies borrow more money when borrowing costs are lower, there is an exemption of Karpavicius and Yu (2017) who deduce that companies do not adjust their capital structures based on interest rates, aside from when they expect a recessionary period. Conversely, interest rates tend to be lower during periods of recession due to interventions by the Central Bank's monetary policy, yet firms lower their demand for external financing. Hence, the relationship can contribute to a positive sign as firms are reluctant to borrow money even though the interest rates are low, since their target ratios are lower.

## RESULTS

Table 1. Mean variables

| YEAR | TOTAL DEBT | LONG DEBT | SHORT DEBT | AS | SIZE | NTC | PROFITAB | NDTS | CASHTA | FINEXP | GROWTH | RISK | LENDING | INFLATION | CREDIT |
|---|---|---|---|---|---|---|---|---|---|---|---|---|---|---|---|
| 1970 | 0.2733 | 0.1799 | 0.0932 | 0.0240 | 3.8002 | 1.4857 | 0.1209 | 13.2078 | 0.0468 | 3.3992 | 1.1550 | 0.6125 | 7.9516 | 2.2114 | 70.8748 |
| 1971 | 0.2648 | 0.1848 | 0.0798 | 0.0231 | 3.8211 | 1.6837 | 0.1265 | 13.5701 | 0.0483 | 3.8162 | 1.1585 | 0.6299 | 7.9100 | 5.0777 | 75.0801 |
| 1972 | 0.2617 | 0.1832 | 0.0784 | 0.0197 | 3.9444 | 1.7367 | 0.1418 | 15.0451 | 0.0463 | 3.7318 | 1.5054 | 0.6320 | 5.7233 | 4.3292 | 75.5509 |
| 1973 | 0.2760 | 0.1858 | 0.0903 | 0.0261 | 3.9212 | 1.2940 | 0.1493 | 18.2348 | 0.0442 | 3.7865 | 1.2901 | 0.6229 | 5.2483 | 5.4420 | 80.2943 |
| 1974 | 0.3215 | 0.2097 | 0.1118 | 0.0625 | 3.4267 | 1.7044 | 0.1288 | 16.4546 | 0.0480 | 5.5157 | 1.2709 | 0.5479 | 8.0216 | 8.9830 | 83.1427 |
| 1975 | 0.3231 | 0.2222 | 0.1010 | 0.0698 | 3.4064 | 1.5027 | 0.1224 | 13.7927 | 0.0524 | 6.1333 | 1.1434 | 0.3658 | 10.798 | 9.2614 | 86.1370 |
| 1976 | 0.3136 | 0.2117 | 0.1020 | 0.0788 | 3.4506 | 7.8123 | 0.1284 | 18.1653 | 0.0534 | 15.506 | 1.3360 | 0.2788 | 7.8625 | 5.4889 | 83.5061 |
| 1977 | 0.3458 | 0.2378 | 0.1082 | 0.0781 | 3.5060 | 3.0467 | 0.1214 | 19.2859 | 0.0510 | 9.5546 | 1.4156 | 0.1128 | 6.8400 | 6.2029 | 87.0183 |
| 1978 | 1.3023 | 1.1973 | 0.1044 | 0.0761 | 3.5430 | 7.4994 | 0.1110 | 22.4531 | 0.0467 | 17.187 | 1.6365 | 0.4517 | 6.8241 | 7.0212 | 87.8505 |
| 1979 | 0.3191 | 0.2174 | 0.1020 | 0.0757 | 3.6510 | 2.2571 | 0.1207 | 28.6891 | 0.0438 | 11.242 | 2.3461 | 0.6301 | 9.0566 | 8.2558 | 86.7595 |
| 1980 | 0.3241 | 0.2215 | 0.1029 | 0.2561 | 3.6022 | 2.6549 | 0.0981 | 29.5212 | 0.0464 | 8.3997 | 1.6352 | 0.7987 | 12.665 | 9.0185 | 87.4663 |
| 1981 | 0.3218 | 0.2109 | 0.1113 | 0.2547 | 3.5652 | 6.3296 | 0.0680 | 25.5831 | 0.0446 | 26.097 | 1.4924 | 0.3939 | 15.265 | 9.3362 | 90.8892 |
| 1982 | 0.3282 | 0.2073 | 0.1215 | 0.0777 | 3.4479 | 4.3758 | 0.0294 | 19.5549 | 0.0440 | 21.634 | 1.6593 | -0.1015 | 18.8700 | 6.2037 | 95.5419 |
| 1983 | 0.3040 | 0.1818 | 0.1228 | 0.0745 | 3.4130 | 9.3877 | -0.0158 | 22.8486 | 0.0493 | 52.694 | 1.4674 | -0.1475 | 14.8608 | 3.9483 | 94.6282 |
| 1984 | 0.3431 | 0.2163 | 0.1273 | 0.1140 | 3.4848 | 5.2038 | 0.0276 | 27.8380 | 0.0455 | 10.720 | 2.5949 | 0.1047 | 10.7941 | 3.5482 | 92.0732 |
| 1985 | 0.6782 | 0.1946 | 0.4843 | 0.1035 | 3.4458 | 19.725 | -0.2502 | 23.5906 | 0.0482 | 167.42 | 1.5363 | 0.1994 | 12.0425 | 3.1996 | 90.2594 |
| 1986 | 0.5021 | 0.2007 | 0.3023 | 0.1848 | 3.4426 | 4.4731 | -0.0775 | 21.6925 | 0.0539 | 20.078 | 2.1467 | 6.21E-05 | 9.9333 | 2.0176 | 89.6848 |
| 1987 | 0.5561 | 0.3618 | 0.2428 | 0.0850 | 3.5384 | 6.3486 | -0.0257 | 31.9560 | 0.0680 | 17.748 | 1.8491 | -0.0162 | 8.3325 | 2.5512 | 92.0733 |
| 1988 | 0.5165 | 0.2123 | 0.3049 | 0.2445 | 3.6599 | 6.1551 | -0.0264 | 41.0806 | 0.0960 | 21.490 | 2.516 | 0.0692 | 8.2033 | 3.5009 | 91.4379 |
| 1989 | 0.5516 | 0.2355 | 0.3172 | 0.3013 | 3.7541 | 6.0851 | -0.0630 | 40.3826 | 0.1056 | 33.118 | 1.5229 | 0.1663 | 9.3150 | 3.8880 | 92.7756 |
| 1990 | 0.5602 | 0.2290 | 0.3324 | 0.1292 | 3.8168 | 5.3501 | -0.0901 | 38.7646 | 0.1010 | 21.230 | 1.6018 | 0.2218 | 10.8733 | 3.6991 | 94.1835 |
| 1991 | 0.5066 | 0.2117 | 0.2958 | 0.2445 | 3.8282 | 8.1652 | -1.5949 | 32.4299 | 0.1131 | 60.567 | 1.5871 | -0.0532 | 10.0091 | 3.3285 | 92.7757 |
| 1992 | 0.4758 | 0.2169 | 0.2594 | 0.1845 | 3.8911 | 6.8181 | -0.0341 | 21.6130 | 0.1116 | 22.4519 | 1.6250 | -0.1369 | 8.4633 | 2.2795 | 92.5776 |
| 1993 | 0.7463 | 0.2147 | 0.5319 | 0.0814 | 3.9534 | 9.3733 | -0.1058 | 28.1765 | 0.1142 | 25.5197 | 1.9161 | -0.2411 | 6.2516 | 2.3792 | 96.0009 |
| 1994 | 0.5489 | 0.1920 | 0.3572 | 0.0626 | 4.0431 | 6.7398 | -0.0730 | 37.4100 | 0.1068 | 36.5553 | 2.5491 | 0.0402 | 6.0000 | 2.1281 | 96.7222 |
| 1995 | 0.5683 | 0.2205 | 0.3475 | 0.0702 | 4.0287 | 5.2205 | -0.1152 | 42.0173 | 0.1229 | 42.2790 | 1.7568 | 0.3021 | 7.1383 | 2.0856 | 103.499 |
| 1996 | 0.5332 | 0.2056 | 0.3277 | 0.0652 | 4.1334 | 8.3853 | -0.0727 | 47.5943 | 0.1354 | 49.0321 | 2.1673 | 0.3983 | 8.8291 | 1.8255 | 109.909 |
| 1997 | 0.3733 | 0.2183 | 0.1550 | 0.0443 | 4.2277 | 9.1648 | -0.1043 | 49.2661 | 0.1370 | 52.4640 | 1.8510 | 0.2650 | 8.2708 | 1.7115 | 112.272 |
| 1998 | 0.4359 | 0.2822 | 0.1536 | 0.0894 | 4.1859 | 7.5559 | -0.2237 | 47.2881 | 0.1428 | 36.7125 | 2.6230 | 0.1245 | 8.4416 | 1.0852 | 113.340 |
| 1999 | 0.5622 | 0.2895 | 0.2728 | 0.2221 | 4.1620 | 10.889 | -0.3662 | 55.1778 | 0.1499 | 73.3415 | 2.3968 | 0.2763 | 8.3541 | 1.5303 | 117.205 |
| 2000 | 0.7126 | 0.3198 | 0.3925 | 0.3906 | 4.2519 | 15.445 | -0.6946 | 59.9323 | 0.1441 | 145.214 | 2.6266 | 0.7787 | 7.9941 | 2.2755 | 114.475 |
| 2001 | 1.5760 | 0.7318 | 0.8417 | 0.2708 | 4.2442 | 9.7609 | -1.3591 | 19.8571 | 0.1489 | 123.189 | 5.4062 | 1.5531 | 9.2333 | 2.2789 | 118.870 |
| 2002 | 1.5664 | 0.3654 | 1.1994 | 1.0743 | 4.3220 | 10.294 | -6.6964 | 34.8389 | 0.1510 | 175.184 | 4.3980 | 0.3447 | 6.9216 | 1.5351 | 117.900 |
| 2003 | 1.4785 | 0.3680 | 1.1092 | 0.7053 | 4.4185 | 11.644 | -1.3886 | 71.2265 | 0.1698 | 90.7721 | 3.8692 | 0.1645 | 4.6750 | 1.9940 | 120.639 |
| 2004 | 2.1433 | 0.3989 | 1.7413 | 0.3421 | 4.5202 | 27.207 | -1.6587 | 106.482 | 0.1707 | 123.404 | 2.3613 | -0.0049 | 4.1225 | 2.7497 | 119.829 |
| 2005 | 1.1282 | 0.2394 | 0.8892 | 0.3038 | 4.6066 | 25.797 | -1.1897 | 131.230 | 0.1726 | 353.7810 | 4.1964 | 0.9996 | 4.3400 | 3.2176 | 129.755 |
| 2006 | 1.2819 | 0.4569 | 0.8231 | 0.0796 | 4.6874 | 17.539 | -1.0460 | 153.125 | 0.1741 | 152.2459 | 2.9589 | 0.8642 | 6.1891 | 3.0722 | 137.205 |
| 2007 | 1.5291 | 0.2194 | 1.3068 | 0.2752 | 4.8022 | 30.474 | -1.3053 | 157.709 | 0.1774 | 202.1193 | 4.0999 | 0.6580 | 7.9575 | 2.6613 | 146.052 |
| 2008 | 1.3820 | 0.3605 | 1.0184 | 0.1653 | 4.9324 | 31.440 | -1.4273 | 97.8908 | 0.1708 | 178.4369 | 1.4921 | 0.2976 | 8.0500 | 1.9616 | 157.347 |
| 2009 | 2.1161 | 0.3286 | 1.7818 | 0.3673 | 4.8376 | 26.190 | -1.4767 | 124.211 | 0.1872 | 114.3571 | 1.9885 | 0.4211 | 5.0875 | 0.7594 | 171.091 |
| 2010 | 1.7777 | 0.5079 | 1.2669 | 0.2528 | 4.9236 | 21.880 | -1.8476 | 155.692 | 0.1963 | 227.0718 | 1.8098 | 0.0700 | 3.2500 | 1.2213 | 162.085 |
| 2011 | 1.1052 | 0.2419 | 0.8613 | 0.2152 | 4.9857 | 32.567 | -2.7259 | 159.685 | 0.1936 | 286.6734 | 1.8782 | 0.2261 | 3.2500 | 2.0646 | 157.348 |
| 2012 | 1.9496 | 0.2464 | 1.6981 | 0.3073 | 4.8051 | 34.226 | -1.7173 | 154.554 | 0.1935 | 300.9128 | 2.0735 | -0.1039 | 3.2500 | 1.8420 | 161.688 |
| 2013 | 1.8299 | 0.7767 | 1.0503 | 0.6484 | 4.8462 | 40.148 | -2.3828 | 158.717 | 0.2007 | 269.4586 | 1.6336 | 0.0471 | 3.2500 | 1.6150 | 176.564 |
| 2014 | 1.4322 | 0.3291 | 1.1008 | 0.1414 | 5.1101 | 39.367 | -2.2605 | 160.511 | 0.1926 | 346.3360 | 4.8132 | -0.2116 | 3.2500 | 1.7903 | 183.936 |

Table 1 contains the timeless process of the mean value of the variables listed in descriptive statistics table. As can be seen the profitability shows that the businesses are not stable and do not have the capacity, even before the crisis, to produce earnings on their spending. The negative average that appears from the beginning (1983) just shows the weaknesses and not the business expenditure audited. The total leverage fluctuated at low levels from 1970 to 2000 and the highest values noticed from 2004 to 2012. The size and growth of US firms remains constant during all the years except a rise in size from 2007 in growth. The cashta variable illustrates that the asset transactions do not affect the prices, which remain unaffected. The tax shield non-interest seems to be used by US companies to reduce their taxes owed. This happens all over the years from 1982 to 2014.

Table2.Correlation of variables

| Correlation | TDRNEW | STDR | LTDR | AS | SIZE | NTC | PROFITABILI... | NDTS | CASHTA | FINEXP | GR | RISK | LENDING | INFLATION | CREDIT SU... |
|---|---|---|---|---|---|---|---|---|---|---|---|---|---|---|---|
| TDRNEW | 1.000000 | | | | | | | | | | | | | | |
| STDR | 0.879650 | 1.000000 | | | | | | | | | | | | | |
| LTDR | 0.511664 | 0.041438 | 1.000000 | | | | | | | | | | | | |
| AS | 0.062453 | 0.070909 | 0.003102 | 1.000000 | | | | | | | | | | | |
| SIZE | -0.042418 | -0.042097 | -0.013061 | -0.066045 | 1.000000 | | | | | | | | | | |
| NTC | -0.000315 | -0.000315 | -9.27E-05 | -0.000141 | -0.078876 | 1.000000 | | | | | | | | | |
| PROFITABILITY | -0.079363 | -0.059993 | -0.058343 | -0.012148 | 0.013857 | -6.91E-05 | 1.000000 | | | | | | | | |
| NDTS | -0.001925 | -0.001660 | -0.001045 | -0.002592 | 0.215397 | -0.003233 | 0.000936 | 1.000000 | | | | | | | |
| CASHTA | 0.038852 | 0.035763 | 0.017012 | -0.004673 | -0.231644 | 0.030268 | -0.005802 | -0.024802 | 1.000000 | | | | | | |
| FINEXP | 0.150862 | 0.161431 | 0.025301 | 0.024714 | -0.077546 | 0.443132 | -0.007526 | -0.002882 | 0.020230 | 1.000000 | | | | | |
| GR | -0.000109 | -1.62E-05 | -0.000200 | -2.89E-06 | -0.011557 | -0.000438 | -0.000149 | -0.001863 | 0.009473 | -0.000476 | 1.000000 | | | | |
| RISK | 0.000418 | -8.58E-05 | 0.001033 | -0.003450 | 0.026352 | 0.002902 | -0.001826 | -0.005010 | 0.026369 | 0.003142 | 0.007645 | 1.000000 | | | |
| LENDING | -0.010041 | -0.009003 | -0.004829 | 0.001615 | -0.165660 | -0.011009 | 0.002340 | -0.058918 | -0.197191 | -0.016741 | -0.003486 | 0.045880 | 1.000000 | | |
| INFLATION | -0.007049 | -0.006811 | -0.002505 | 0.000277 | -0.138306 | -0.009550 | 0.003932 | -0.038467 | -0.226989 | -0.011937 | -0.006622 | 0.181078 | 0.480065 | 1.000000 | |
| CREDIT_SUPPLY | 0.013903 | 0.011989 | 0.007548 | 0.000125 | 0.217258 | 0.015339 | -0.002738 | 0.085751 | 0.254927 | 0.024590 | 0.004421 | 0.022767 | -0.604067 | -0.558608 | 1.000000 |

Table 2 illustrates the correlation coefficients between the variables used in our model. The dependent and independent variables are provided with a Pearson correlation matrix. It is conspicuous that there is a non-statistically significant and negative correlation (r=-0.0010) between long-term debt values and non-debt tax shields. Similar findings were obtained by Frank & Goyal (2003), in which there is a statistically significant and positive correlation (r=0.0004) between total debt and risk at the 5% significance level, a statistically significant and negative correlation (r=-00793) between total debt and profitability at the 5% significance level. Pecking-order theory predicts a negative relationship between profitability and total debt. This theory argues that companies will prefer to finance their needs first by using sustainable profits, then through borrowing and finally through the issuance of new shares. According to the Pecking-order theory, companies that are rapidly developed and have high funding needs will move on to a short-term funding that is less subject to asymmetric information.

Moreover, Pearson correlation provides us information about negative relation (r=-0.0424) between total debt and size. This finding is against trade-off theory, namely that the bigger the business is, the greater the ability to borrow and therefore it can have a higher leverage than a smaller company. According to Titman & Wessels (1988), the bigger the companies are, the more diversified the size will be and subsequently the shorter the probability of bankruptcy will be, as well as the less volatility will be observed in their cash flows. In this way, firms are able to borrow from smaller companies. Finally, there is a statistically significant and negative correlation (r=-0.0002) between long-term debt and growth at the 5% significance level in contrast to trade-off theory.

We observe some differences in regressors attitude of our regressors in the two periods. The effect of the firm and macro regressors cannot be easily extracted. Therefore, we are performing hypothesis testing on the individual dummy and we also group these dummies across macro and firm type. Concerning the macroeconomic variables, credit supply has a persistent and relatively strong effect across states and forms of ratios, inflation is significant in most quantiles and interest rates show an interesting shift from relatively negative effect before the crisis to positive effect during the crisis.

**Hausman Test**

With the regression equation, we will choose the most catalyzed model between fixed effects and random effects and with the help of the Hausman test we will make the most appropriate choice for our model.

Correlated Random Effects - Hausman Test
Equation: Untitled
Test cross-section random effects

| Test Summary | Chi-Sq. Statistic | Chi-Sq. d.f. | Prob. |
|---|---|---|---|
| Cross-section random | 2916.273546 | 12 | 0.0000 |

Correlated Random Effects - Hausman Test
Equation: Untitled
Test cross-section random effects

| Test Summary | Chi-Sq. Statistic | Chi-Sq. d.f. | Prob. |
|---|---|---|---|
| Cross-section random | 2916.273546 | 12 | 0.0000 |

Correlated Random Effects - Hausman Test
Equation: Untitled
Test cross-section random effects

| Test Summary | Chi-Sq. Statistic | Chi-Sq. d.f. | Prob. |
|---|---|---|---|
| Cross-section random | 2047.762746 | 12 | 0.0000 |

$H_0$ : Random effects model is appropriate

$H_1$ : Fixed effect model is appropriate.

Probability of Chi-Sq < 0.05, so we reject null hypothesis, and Fixed Effect Model is the most appropriate for our model.

We focus our analysis on the relative importance and the subsequent differences between firm-specific and macroeconomic variables, between the macroeconomic states of growth and recession and across the three forms of leverage. We thus perform the hypothesis testing on the significance of the multiplicative dummies. Specifically, we look at the significance of each coefficient,

$H_0 : \delta^C = 0$, $H_0 : \beta^c k = 0$, k=1,...,K, $H_0 : \gamma^c j = 0$, j=1,...,J.

On hypothesis, results show that the hypothesis $H_0 : \delta^C = 0$ is rejected for LTDR only in 2th quantile indicates that firms with long terms debt in low values slows down their adjustment speed during crisis. The hypothesis is not rejected for all the rest quantiles. This hypothesis is rejected for STDR in the most quantiles, meaning that the adjustment speed does not change for STDR during crisis. Specifically, the adjustment speed for STDR for almost all quantiles is higher in the recessionary period but not significantly different from the growth period, whereas speed adjustment of LTDR significantly slows down during crisis. Specifically, and as discussed before, the adjustment speed for STDR is higher in the recessionary period but not significantly different from the growth period, whereas speed adjustment of the LTDR significantly slows down during crisis. Regarding TDR, the result for the adjustment speed is mainly driven by that for STDR; this is expected given that on average STDR forms the bigger part of TDR. Concerning the hypothesis $H_0 : \beta^c k=0$, and $H_0 : \gamma^c j=0$, we reject for k=AS in the first quantile, for size in 7 and 8 quantile, for NTCS in 1 and 9 quantile, for PROF in 4 quantile, for NTDS in 9 quantile, for GR and risk in most of quantiles, and for j= INTR in 3 quantile, INFL in 6 quantile and CREDIT SUPPLY for 5, 6 and 9 quantile for STDR. We reject for k= SIZE, NDTS, CASHTA, GR, RISK in most quantiles and for j= INTR, INFL, CREDIT SUPPLY for LTDR.

**CONCLUSION**

The first thing we find is that the lagged value of the debt ratio is the main determinant across all three forms of leverage, namely short-term, long-term and total, and therefore the remaining firm and macroeconomic factors can play a secondary role in their dynamic determination. Moreover, long-term and short-term debt ratios as regards their adjustment speeds follow different patterns, having the adjustment speed for LTDR slow down during the crisis, whereas the one of STDR is not affected at all. Furthermore, we reach a conclusion that between the two states there is a clear differentiation of the contribution and the effects of the firm-specific vs. the macroeconomic variables for STDR and LTDR. More particularly, macro variables increase their significance in the recessionary period, while firm variables are more significant than macro variables in determining STDR in the growth period variables. Regarding LTDR, macro variables are more significant in determining the ratio in the whole sample period than firm variables, and they become even more significant during the crisis. Due to these explicit differentiations between the two forms of debt, there is definitely no point in drawing inferences on the total debt ratio and thus we drop our conclusions on TDR. Since our conclusions remain invariable, we can deduce that they are robust to most of the checks. The only change that is persistent is indicated for STDR, insinuating that the contribution of firm variables to the mean could be even less significant during the crisis.

Hence, we affirm that the maturity and nature of borrowing itself influences both the endurance and persistence of the relation between determinants and borrowing, in compliance with the general approach of differentiating between long-term and short-term debt for SMEs (Koeter-Kant and Hernandez-Canovas, 2011; García-Teruel and Martínez-Solano, 2010; Michaelas et al., 1999). Considering a case that the firm-specific factors are internal to the firm meaning and managers have a determinate level of control upon these factors, whilst there is no such control over the macroeconomic variables, our results denote that SMEs' managers have very low levels of flexibility for altering the capital structure of the firms they manage during the crisis. In plain words, during the crisis SMEs are particularly vulnerable on how their capital structure is being determined.

It is indisputably implied from our study that the capital structure puzzle, that was properly referred to as such by Myers (1984), which is a multi-dimensional riddle, is simultaneously influenced by the inherent specificities of the following contrary pairs: (a) growth vs. recessionary states, (b) large enterprises vs. SMEs, (c) countries' specificities vs. similarities and (d) short-term vs. long-term debt. Concerning the practical implications of our study, viewed by a managerial perspective, a cumbersome capital structure can be seen as a hindrance in value maximization. Viewed from a regulatory perspective, it is implied that regulators should provide the context within which a financial environment is capable to exist, where better institutions will lead to higher adjustment speed of capital structure as well as lower transaction costs. To that end, compared to large enterprises, this flexible financial environment is of higher significance for SMEs, since the former are comparatively more restricted in access to finance and hence more vulnerable in macroeconomic state changes.

Nevertheless, a main constraint in our study is that our results apply to the specificities of one particular country, and therefore conclusions are restricted to economies that share similar characteristics with America. In conjunction with this, the persistence and depth of the economic recession in America may seem as an extreme environment that is doubtful to be observed in developed economies in the near future. On the contrary, this specific environment is possibly the ideal context to test and experiment with this particular problem in question, considering that the two individual macroeconomic states are distinctively different. A plausible future research could investigate whether capital structure determinants behave differently in economies with milder differences in macroeconomic states and different levels of financial integration..